\documentclass[aps,twocolumn,showpacs,preprintnumbers,amsmath,amssymb]{revtex4}
\usepackage{epsfig}	
\usepackage{dcolumn}    
\usepackage{bm}         
\usepackage{color}      

\begin{document}

\title{Antispiral waves are sources in oscillatory reaction-diffusion media
\footnote{submitted to J. Phys. Chem. B on Feb. 20, 2004.}}


\author{Ernesto M. Nicola$^{a,b,c}$, 
        Lutz Brusch$^{a,d}$
	and Markus B\"ar$^{a}$}
\email{baer@mpipks-dresden.mpg.de}
\affiliation{
$^{a}$ Max-Planck-Institut f\"ur Physik komplexer Systeme,
N\"othnitzer Stra{\ss}e 38, D-01187 Dresden, Germany\\
$^{b}$ Dept. E.C.M., Facultat de F{\'{\i}}sica, Universitat de
Barcelona, 
Av. Diagonal 647, E-08028 Barcelona, Spain \\
$^{c}$ Institut f\"ur Theoretische Physik,~Technische Universit\"at
Berlin, PN~7-1, 
Hardenbergstra{\ss}e 36, D-10623 Berlin, Germany\\
$^{d}$ Centre de Bioing{\'e}nierie Gilbert Durand INSA-DGBA,  
Av. de Rangueil 135, F-31077 Toulouse Cedex 4, France 
}

\begin{abstract}
\noindent
Spiral and antispiral waves are studied  numerically in two examples of
oscillatory reaction-diffusion media and analytically 
in the corresponding complex Ginzburg-Landau equation (CGLE).
We argue that both these structures are sources of waves in
oscillatory media, which are distinguished only by the sign of the
phase velocity of the emitted waves. 
Using known analytical results in the CGLE,  we obtain a criterion for
the CGLE coefficients that  predicts whether antispirals or spirals
will occur in the corresponding reaction-diffusion systems. 
We apply this criterion to the FitzHugh-Nagumo and Brusselator models
by deriving the CGLE near the Hopf bifurcations of the respective
equations. 
Numerical simulations of the full reaction-diffusion equations confirm
the validity of our simple criterion near the onset of oscillations. 
They also  reveal that antispirals often occur near the onset and turn 
into spirals further away from it.
The transition from antispirals to spirals is characterized by a
divergence in the wavelength. 
A tentative interpretaion of recent experimental observations of 
antispiral waves in the Belousov-Zhabotinsky reaction in a microemulsion 
is given. 
\end{abstract}

\maketitle

\section{Introduction}
Chemical pattern formation results typically from the interplay of
reaction and diffusion and occurs in many reactions in solutions and gels 
\cite{Showalter} and in heterogeneous catalysis \cite{ImbErtl}.
Reaction-diffusion processes are also believed to be at the heart of
morphogenesis in biological systems \cite{Murray,keener}. 
Rotating spiral waves are probably the most typical structures 
investigated so far. 

They have been found initially in the Belousov-Zhabotinsky 
reaction \cite{winfree72}.
Since then, they have been  frequently observed in a
variety of experimental setups including aggregation of slime molds
\cite{dicty}, catalytic CO oxidation  \cite{catalysis}, cardiac tissue 
\cite{jalife}, and intracellular calcium dynamics in frog eggs \cite{lechl}, 
and glycolytic activity in extracts of yeast cells \cite{mair}.
Spirals that organize the surrounding medium by regular emission of waves
are well established under both oscillatory and excitable conditions.
If several spirals populate the system then the waves emitted by neighbouring
spirals annihilate each other at the boundary of the spirals's spatial domains.
Such boundaries constitute wave sinks.
Intuition tell us that these passive objects (where waves arrive and die) do not
influence the medium beyond its closest surrounding.

Recently, Vanag and Epstein 
reported on experiments 
of the oscillatory Belousov-Zhabotinsky reaction in an Aerosol OT
microemulsion (BZ-AOT)~\cite{Vanag}.
In their quasi two-dimensional system chemical concentration patterns are 
arranged as so-called ``antispirals'' or ``antitargets''. 
Typical snapshots of the system display closely packed domains each containing 
a single antispiral. 
As time progresses the concentration waves propagate {\it towards} the center 
of the respective antispiral. 
Such inward propagation seemingly contradicts our intuition of spiral
cores as wave sources. 
A word of caution towards the terminology of spirals and antispirals
is in place here:  
Spirals and antispirals are not structures that can appear
simultaneously for given parameters. 
Instead, they occur in different complementary regions of the
parameter space. 
The terminology of antispirals used here is rather recent and occurs
only after the cited experimental paper, that is more careful and talks of
inwardly rotating spiral waves. 
It should also be stressed that many authors have found (what is now
often called)  antispirals or antitargets in numerical simulations in continuous 
and discrete oscillatory and chaotic media and simply classified them 
as spirals or targets see {\it e.g.}
\cite{YK76,selpuchre,kapral,Ipsen97,TK98,Rabinovitch,Stich}. 
More recent reports have adapted the new name antispiral in the sense
it is used here  \cite{Lee,soerensen}.
The distinction achieved here by the prefix anti  is sometimes also
referred to as positive  dispersion (spiral) and negative dispersion
(antispiral) \cite{Stich}. 
Since, the phenomena of antispirals and antitargets occur 
frequently in the complex Ginzburg-Landau 
equation  \cite{review}, they
should be generic in oscillatory media near onset. 
What comes as a surprise is rather that over so many years only
spirals have been observed and the first experimental report of an 
antispiral occured only very recently \cite{Vanag}. 
This experiment lead to a discussion about the ``mechanism'' of
antispirals. 
Vanag and Epstein suggested that the phenomenon may be related to a
Hopf bifurcation with finite wavenumber (wave bifurcation) in a
extended Oregonator model, where they find
group and phase velocities with opposite signs. 
Simulations and experiments show also ``packet waves'', which are 
wave groups that move in opposite direction to the motion of the 
constituting waves \cite{Vanag3}. 
While this argument seems valid for the quasi one-dimensional packet
waves, it does, in our opinion, not apply to the two-dimensional case
under consideration. 
After all, near a wave bifurcation in a two-dimensional medium, one
typically observes traveling or standing stripes or hexagons and not 
spiral waves.
Earlier on, Nicola {\it  et al.}  have reported related  wave groups near a
Turing-wave bifurcation and named them ``drifting pattern domains''
\cite{Nicola}. 

Gong and Christini have recently investigated the CGLE
and prototypical two-component reaction-diffusion (RD) models and
conjectured that antispirals  only occur near the onset of oscillations \cite{christini}.
We will provide further evidence for their claim. 
They also suggest  an analytical argument and criterion for the
appearance of antispirals in the CGLE coefficients. 
Recently, we have shown that the criterion for the occurrence of
antispirals in RD models has to be modified and that antispirals in the
corresponding CGLE may turn out to be spirals in the original RD model
\cite{BNB03}.
Here we present the complete analytical derivation of the differing  
criteria  for antispirals in the CGLE and in the corresponding RD
models. 

When studying dynamical phenomena in RD models
near the onset of pattern formation then the analysis of the corresponding 
amplitude equation can be  instructive \cite{Kuramoto,CH,Nicolis}.
The CGLE represents the amplitude equation for pattern dynamics  
near the onset of oscillations via supercritical
Hopf bifurcations \cite{review}.
The CGLE  for arbitrary RD systems is obtained following exact
transformation rules \cite{Ipsen97,Ipsen00}.
Here, we validate our criterion for antispirals in RD systems 
for two simple reaction-diffusion (RD) systems: 
the FitzHugh-Nagumo and  Brusselator models by deriving the CGLE 
coefficients for these models, applying our criterion for antispirals 
and comparing the result to direct numerical simulations near a 
Hopf bifurcation. 
In the CGLE, domains of the parameter space can be classified
according to the relative directions (signs) of phase- and
group-velocities in rotating waves (spirals or antispirals)
 in line with arguments which 
Y. Kuramoto and T. Tsuzuki used first to analyse wave 
sources of the Kuramoto-Sivashinsky equation \cite{KT76}.
Boundaries of the parameter domains are given by zero phase
 or group velocities which yields analytical criteria.
Extensive numerical simulations of both  RD models analysed here, 
corroborate this criterion's prediction near the onset of oscillations
and extend it when the model's parameters are driven away form threshold. 
The simulations also suggest that antispirals typically disappear far away from
the onset of oscillations. 

This paper is organised as follows: 
In the next section we will discuss theoretically
the distinction between spirals and antispirals within the CGLE framework and
analytically derive the criterion for either occurrence in RD models near the
onset of oscillations.
In the third section we explore spirals and antispirals in two 
RD models near and far from the onset of oscillations.
We will end this paper with a short summary and discussion of the main results. 
%

\section{Weakly nonlinear theory of spirals and antispirals}

\subsection{RD systems near a oscillatory bifurcation}

\begin{figure*}
  \begin{center}
    \epsfxsize=0.9\hsize \mbox{\hspace*{-.06 \hsize} \epsffile{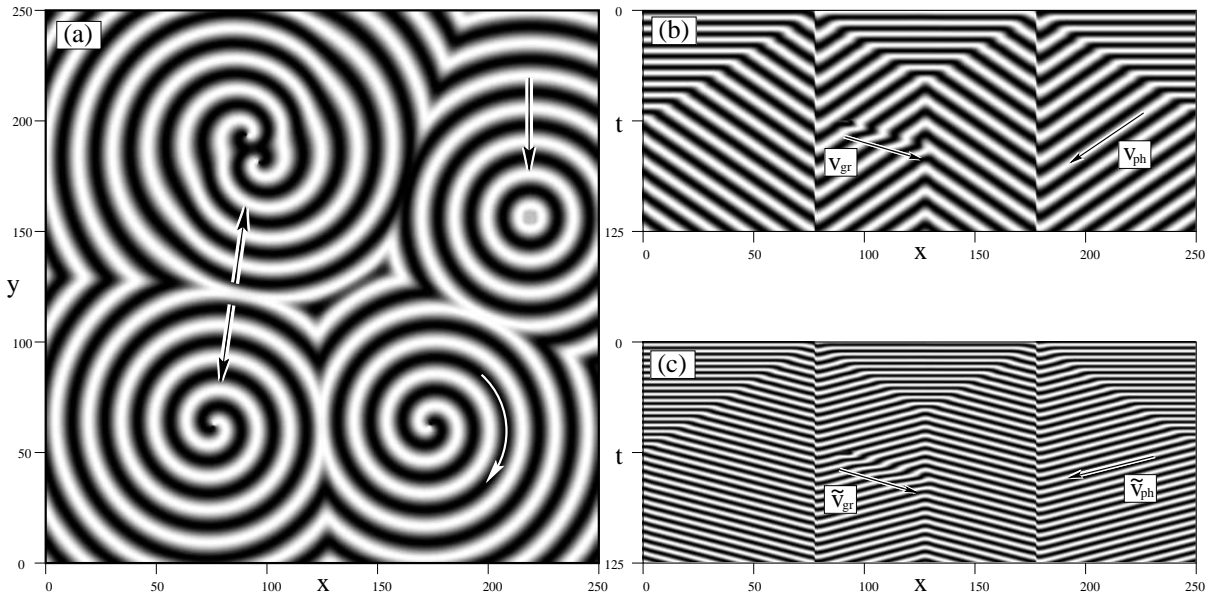} }
  \end{center}
\caption{
Numerical simulation of the CGLE (\ref{eq:cgle})  revealing antispirals and antitargets for $c_1=1, c_3=0.5$.
In (a)  the real part of $A$, in a system of size 250*250 with zero flux boundary 
conditions and after a transient of $t=125$, is shown. 
White (black) areas correspond to maximum (minimum) values.
The black arrows indicate the direction of propagation of phase waves and the white arrow 
denotes the rotation of an individual spiral. 
The inwardly propagating target waves are induced by a small oscillating heterogeneity in their center. 
The domain boundaries, where new waves emerge and split, are visible between the 
antispirals.
In (b) a space-time plot of $Re[A]$, along a horizontal cut at $y=60$
(this line contains the cores of the antispirals at $x=75$ and $175$), is shown.
After a transient of $t=125$ the final state leads to the configuration shown in (a).
Initially the cores are placed in the homogeneously oscillating background. 
This homogeneous background is suppressed as the antispirals grow until the
whole system is filled by the antispirals at $t=60$.
Note how new waves emerge and split at the domain boundary at $x=125$. 
The evolution of a small perturbation in (b) illustrates the different
sign of the phase and group velocities $v_{ph}$ and $v_{gr}$ 
(the artificial perturbation is applied at $x=90, t=65$ and propagates with $v_{gr}$).
The space-time plot shown in (c) corresponds to the same situation
than in (b) but a fast homogeneous oscillation with $\Omega=0.02$ has been added to the phase of $A$. 
This illustrates the dynamics in the underlying RD system which, for the value of $\Omega$ chosen, also shows 
antispirals ({\it i.e.} the phase velocity $\tilde{v}_{ph}$ and the group velocity
$\tilde{v}_{gr}$ have the same direction as in the CGLE).
}
 \label{fig:snap}
\end{figure*}

A general reaction-diffusion system in two dimensions may be described 
by the partial differential equation
\begin{equation}
\partial_{\tilde{t}} {\bm u} = {\bm f}({\bm u},\mu)+{\cal D} \nabla^2 {\bm u},
\label{eq:RD-system}
\end{equation}
where ${\bm u}(\tilde{\bm x},\tilde{t})$ is a vector of space- and 
time-dependent concentrations, ${\bm f}$ is a nonlinear vectorial function, 
${\cal D}$ is a diffusion tensor and $\mu$ is a control parameter.
Realistic models of chemical pattern forming systems have been proposed in this 
form including recent models of the BZ-AOT system by Vanag {\it et
al.} \cite{Vanag2,Vanag3} and Yang {\it et al.} \cite{Yang}.
These include many reaction species, respectively components of $\bm u$, and 
complex reaction terms $\bm f$. 
However,  in this paper we shall study simple models namely the
Brusselator~\cite{Brusselator} or the FitzHugh-Nagumo
model~\cite{FHN}.
(Anti)spiral waves in the latter two models will be studied in 
Sec.~\ref{sec:rd} and explained by results of the weakly nonlinear analysis
that we carry out in the present section.
Near the supercritical onset of homogeneous oscillations in the RD
system with frequency $\Omega$ and eigenvector ${\bm u}_1$ the vector of concentrations
$\bm u$ may be decomposed as
\begin{equation}
   {\bm u}(\tilde{\bm x},\tilde{t})=
     {\bm u}_0 
   + {\bm u}_1   \tilde{A}  ({\bm x},t) \, \mbox{e}^{i \Omega \tilde{t}} 
   + {\bm u}_1^* \tilde{A}^*({\bm x},t) \, \mbox{e}^{-i \Omega \tilde{t}}   ~.
 \label{eq:ansatz}
\end{equation}
The evolution of modulations 
$\tilde{A}({\bm x},t)= \sqrt{\epsilon} A({\bm x},t) \mbox{e}^{i c_0 t}$ 
of a homogeneous oscillation  is described by the CGLE \cite{CH}
\begin{equation}
   \partial_t A=A+(1+i c_1) \Delta A - (1-i c_3) |A|^2 A ~.
 \label{eq:cgle}
\end{equation}
Here $\epsilon=(\mu-\mu_c)/\mu_c$ measures the distance from the threshold
$\mu_c$, $c_1$ ($c_3$) give the 
linear (nonlinear) dispersion and $c_0$ an overall linear frequency shift.
The coordinates ${\bm x}=\sqrt{\epsilon} \frac{\tilde{\bm x}}{\xi}$ and 
$t=\epsilon \frac{\tilde{t}}{\tau}$ of the CGLE are defined by characteristic
spatial and temporal scales $\xi$, $\tau$ and describe slow modulations in space
and time.
Note, that intrinsic CGLE frequencies $\omega$ result only in a small correction
of order $\epsilon \omega$ to the original frequency of the system at the Hopf bifurcation threshold.
In the following, variables with (without) tilde will be used in the original 
RD system (the derived CGLE).
Numerical simulations of the CGLE (\ref{eq:cgle}) in a two-dimensional system with
zero flux boundary conditions provide examples of antispiral waves as shown in
Fig.~\ref{fig:snap}.
%
%

\subsection{Spirals and antispirals in the CGLE}


%
These (anti)spiral waves of the two-dimensional CGLE have the 
form \cite{Hagan}
\begin{equation}
   A(r,\theta,t) = F(r) \mbox{e}^{\displaystyle i(\sigma\theta+f(r,t))} ~,
 \label{eq:spiralform}
\end{equation}
in polar coordinates ($r,\theta$). 
The asymptotic behaviour for $r\to0$ is 
$\partial f(r,t)/\partial r \sim r, F(r) \sim r$
and denotes a topological defect $|A|=0$ of charge $\sigma$ in the spiral core.
We will focus on one-armed spiral waves with $\sigma=+1 (-1)$ which 
are related by mirror symmetry and rotate clockwise (counterclockwise).
Other authors term these solutions spiral and antispiral \cite{anti1} (we will not adopt this terminology here). 
In this paper, both left-handed or right-handed structures can be
spirals (outward propagation of phase 
waves) or antispirals (inward propagation of  phase waves) depending on parameters.

Asymptotically for $r\to\infty$ these phase waves are characterised by
$F(r) \sim \sqrt{1-q_s^2}$ and $f(r,t) \sim q_s r-\omega_s(q_s) t$
with a selected wavenumber $q_s$ and the corresponding frequency $\omega_s(q_s) = -c_3+q_s^2(c_1+c_3)$.
The selected wavenumber $q_s$ is a function of the parameters $c_1$ and $c_3$ given by the nonlinear eigenvalue problem resulting form inserting Eq.~(\ref{eq:spiralform}) into the CGLE (\ref{eq:cgle}) \cite{Hagan}.
The asymptotic wave field $f(r,t)$ may be rewritten as $q_s (r - v_{ph} t)$
with the phase velocity $v_{ph}=\omega_s / q_s$~.
The propagation of small perturbations of the wave is described by the 
group velocity
$v_{gr}=\partial \omega_s / \partial q_s = 2 q_s (c_1+c_3)$~.
For positive group velocity the spiral or antispiral acts as a source that
organises the surrounding pattern.
Phase and group velocity do not necessarily point into the same direction.
The phase waves move outward (inward) for positive (negative) 
phase velocity $v_{ph}$ in the CGLE coordinates or $\tilde{v}_{ph}$ in the 
RD system, respectively.
Hence, the signs of the selected group and phase 
velocities constitute the defining quantities for the occurrence of
spirals or antispirals.

%
In the following we calculate the parameter dependence of the introduced
velocities and focus on their signs.
The selected wavenumber $q_s (c_1, c_3)$ needs to be computed numerically in 
general, only its asymptotic are known analytically \cite{Hagan}.
However, for the one-dimensional (1D) analog of the (anti)spiral wave ({\it i.e.} for the 1D CGLE in $[0,\infty)$ with Dirichlet boundary condition: $A(0,t)=0$) the corresponding selected wavenumber $q_{s1} (c_1, c_3)$ 
is  known analytically \cite{Hagan,q1D}:
\begin{widetext}
\begin{subequations}
\label{eq:qhagan1D}
\begin{eqnarray}
 q_{s1} &=& - \frac{3 \alpha(c_1,c_3)}{2 (c_1 +c_3)} {+ \atop -}
   \sqrt{\frac{9 \alpha(c_1,c_3)^2}{4 (c_1 +c_3)^2}+
   \frac{c_3+2 c_1 \alpha(c_1,c_3)^2}{c_1+c_3}},
\quad \quad \mbox{for} ~c_1 + c_3~ {> \atop <} ~0 ~,\\
\mbox{where} \quad \quad
\alpha(c_1,c_3) &=& 
\sqrt{\frac{3 c_1 (8 (c_1-c_3)^2+9 (1+c_1 c_3)^2-4 c_1 c_3)^{1/2}
  +c_1 (5-9 c_1 c_3) -4 c_3}{4 (-2 c_3+9 c_1^3+7 c_1)}} ~.
\end{eqnarray}
\end{subequations}
\end{widetext}
The function $\alpha(c_1,c_3)$ is symmetric and $q_{s1} (c_1,c_3)$ anti-symmetric 
under the substitution $(c_1 \to -c_1, c_3 \to -c_3)$.
The same property holds for the two-dimensional (anti)spirals as can be deduced from the
form of the nonlinear eigenvalue problem \cite{Hagan}:
\begin{subequations}
\begin{eqnarray}
q_s (c_1, c_3) &=& -q_s (-c_1, -c_3),  \label{subeq:qsym} \\
q_s (c_1, c_3) &=& 0 \quad,\quad\mbox{for} ~c_1+c_3 = 0,  \label{subeq:qsym2} \\
q_s (c_1, c_3) &>& 0 \quad,\quad\mbox{for} ~c_1+c_3 > 0,  \label{subeq:qsym3} \\
q_s (c_1, c_3) &<& 0 \quad,\quad\mbox{for} ~c_1+c_3 < 0.  \label{subeq:qsym4}
\end{eqnarray}
\end{subequations}
Hence, $q_s$ takes the same sign as $c_1+c_3$.
%
\begin{figure}[t]
\begin{center}
 \epsfxsize=0.98\hsize \mbox{\epsffile{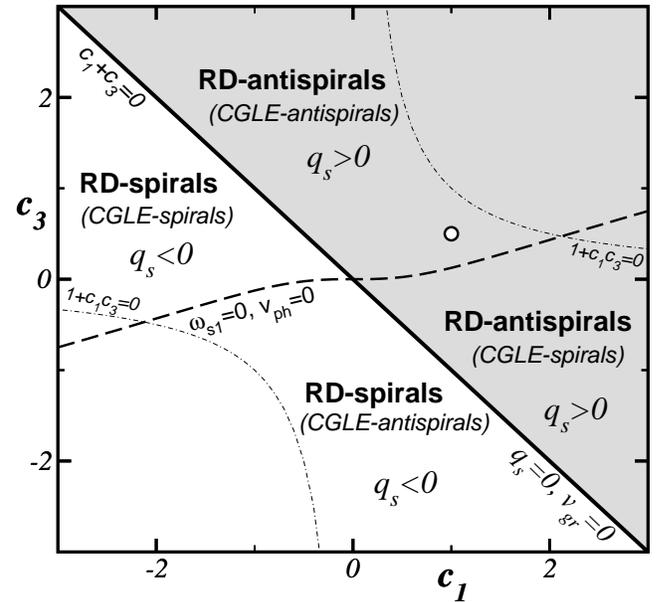} }
\end{center}
 \caption{
 Parameter space $(c_1, c_3)$ of the complex Ginzburg-Landau equation (CGLE) 
 and domains where spiral or antispiral wave solutions exist in the
 corresponding RD systems (where we assumed that $\Omega>0$). 
 The group velocity of spirals and antispirals is non-negative everywhere.
 Note that in the CGLE the phase velocity vanishes along the dashed curve and antispirals
 (spirals) occur in the upper (lower) shaded and lower (upper) white regions.
 The dot-dashed curves denote the Benjamin-Feir instability $(1+c_1 c_3=0)$.
The white circle indicates the location of parameter values used in Fig. \ref{fig:snap}.
 }
 \label{fig:c1c3}
\end{figure}
%
The corresponding frequency $\omega_s (c_1, c_3)$ in the CGLE fulfils
$\omega_s (c_1, c_3) = -\omega_s (-c_1, -c_3)$ as had already been noted by Paullet {\it et al.}~\cite{Ermentrout}.
The curve where the selected frequency $\omega_{s1}$ vanishes has been computed numerically in the $(c_1, c_3)$ parameter space ({\it i.e.} we computed the parameters $c_1$  and $c_3$ for which $\omega_{s1} (q_{s1}(c_1, c_3)) = 0$).
It turns out to be different from the diagonal $c_1 + c_3 = 0$ where $q_s=0$. 
Hence, within the CGLE the parameter space, we can distinguish four different domains,
as shown in Fig.~\ref{fig:c1c3}. 
When crossing the boundaries between these domains, the phase velocity $v_{ph}$ in the CGLE switches sign between outward and inward.
Note, that the spiral wave length and $v_{ph}$ diverge on the diagonal and
$v_{ph}$ crosses zero where $\omega_s=0$.
In the coordinate frame of the CGLE, there are two domains with spirals and two with antispirals.
This result is in agreement with that of Gong {\it et al.}   \cite{christini}
and additionally covers two quadrants previously not considered, 
where the transition curve $v_{ph}=0$ lies.
Inserting Eq.~(\ref{subeq:qsym}) into $v_{gr}=2 q_s (c_1+c_3)$ one finds that the group velocities $v_{gr}$ of both spirals and antispirals never become negative.

\subsection{Spirals and antispirals in RD systems}

%
Now we return to the initial RD system.
Therein the asymptotic concentration waves 
$\bm u \sim \mbox{e}^{i \tilde{q}_s (\tilde{r}-\tilde{v}_{ph} \tilde{t})}$
with wavenumber 
$\tilde{q}_s=\sqrt{\epsilon} q_s /\xi$ read, after inserting 
Eq.~(\ref{eq:spiralform}) into Eq.~(\ref{eq:ansatz}),
\begin{eqnarray}
  {\bm u}(\tilde{r},\tilde{\theta},\tilde{t}) &=&
  {\bm u}_0 + {\bm u}_1
\sqrt{\epsilon} \sqrt{1-\frac{\xi^2 \tilde{q}_s^2}{\epsilon}} ~
\times \nonumber\\
& &\mbox{e}^{\displaystyle i[\sigma\tilde{\theta}+ 
\tilde{q}_s \tilde{r} - (-\Omega+\epsilon(\omega_s-c_0)) \tilde{t}]} \nonumber\\
& &+ \mbox{c.c.} \label{eq:wave1} ~,
\end{eqnarray}
which yields
\begin{eqnarray}
\tilde{v}_{ph}&=&\left[-\Omega+\frac{\epsilon}{\tau} (\omega_s-c_0) \right] /
                 \tilde{q}_s ~. 
\label{eq:wave2}
\end{eqnarray}

The description given by the CGLE (\ref{eq:cgle}) is accurate only if $\epsilon \ll 1$. 
In this limit
 $\tilde{v}_{ph}$ may be simplified as
$\tilde{v}_{ph} \approx - \Omega/\tilde{q}_s \sim - \Omega/q_s$~
(note that the same result follows for the complex conjugate (c.c.) in 
Eq.~(\ref{eq:wave1}), where two negative signs cancel).
For the group velocity we find 
$\tilde{v}_{gr}=v_{gr} \frac{\sqrt{\epsilon} \xi}{\tau} \sim v_{gr}$
which does never get negative.
Therefore, both spirals and antispirals will act as organising centers 
of the surrounding concentration pattern.

In order to determine the parameter dependence of the phase velocity 
$\tilde{v}_{ph}\sim - \Omega/q_s$, we first assume $\Omega > 0$. 
Then spirals with positive phase velocity occur in RD systems with
corresponding $c_1+c_3 < 0$ as follows from Eq.~(\ref{subeq:qsym3}).
Antispirals have $q_s > 0$ and arise in RD systems for which $c_1+c_3 > 0$.
Consequently, spirals and antispirals can never occur simultaneously within a single 
homogeneous system.
These analytical results are summarised in Fig.~\ref{fig:c1c3} (where $\Omega > 0$ 
was chosen).
Let us now briefly comment on the case of $\Omega <0$ 
(note that  the remarks that follow are not important since by
convention a $\Omega >0$ is considered in order to calculate the amplitude equation (\ref{eq:cgle})).
Assuming the opposite sign of the primary oscillation $\Omega \to -\Omega$,
yields the same results for the group and phase velocities.
For negative $\Omega$ we find
$c_1 \to -c_1, c_3 \to -c_3, \omega \to -\omega$
as follows from complex conjugation of Eq.~(\ref{eq:cgle})
and $q_s \to -q_s$ as in Eq.~(\ref{subeq:qsym}).
This gives the condition $c_1+c_3 < 0$ for the existence of antispirals if  $\Omega <0$.
Using Eqs.~(\ref{eq:wave2}) and (\ref{subeq:qsym3}), we compare the temporal frequencies 
$f=-\tilde{q}_s \tilde{v}_{ph}$ measured for a spiral ($f_{\mbox{\small S}}$) or 
antispiral ($f_{\mbox{\small AS}}$) wave with  that of the homogeneous 
oscillation $f_{\mbox{\small bulk}}$
\begin{subequations}
 \begin{eqnarray}
  f_{\mbox{\small bulk}} &=& \Omega + \frac{\epsilon}{\tau} (c_0+c_3) ~,\\
  f_{\mbox{\small S}} &=& \Omega + \frac{\epsilon}{\tau} (c_0+c_3) +
               \frac{\epsilon}{\tau} q_s^2 |c_1+c_3| ~,\\
  f_{\mbox{\small AS}}&=& \Omega + \frac{\epsilon}{\tau} (c_0+c_3) - 
               \frac{\epsilon}{\tau} q_s^2 |c_1+c_3|   ~.
 \end{eqnarray}
\end{subequations}
We conclude 
$f_{\mbox{\small S}} > f_{\mbox{\small bulk}}$ and
$f_{\mbox{\small AS}}< f_{\mbox{\small bulk}}$
in agreement with experimental observations \cite{Vanag}.

Let us finish this section with a short summary.
The observed antispiral waves with phase velocity pointing inward were shown 
to be organising centers since their group velocity points outward. 
Antispirals in oscillatory media therefore determine the surrounding pattern as 
it is the case for spiral waves.
We found that a spiral in the CGLE coordinate frame may represent an antispiral
in the original RD system and vice versa (see Fig.~\ref{fig:c1c3}).
Altogether, in RD systems near the onset of oscillations, 
the antispirals are predicted to occur if the corresponding CGLE coefficients 
fulfil $c_1+c_3 > 0$.
%

\section{Reaction-diffusion models}
\label{sec:rd}

In order to test the general predictions of the previous section 
we will now address our attention to a specific RD systems.
The RD systems of activator-inhibitor type are defined by \cite{Nicolis} 
\begin{subequations}
 \label{eq:AI}
 \begin{eqnarray}
    \partial_{\tilde{t}} u &=& f(u,v) + \nabla^2 u ~,      \label{subeq:AI-u}\\
    \partial_{\tilde{t}} v &=& g(u,v) + \delta \nabla^2 v ~,\label{subeq:AI-v}
 \end{eqnarray}
\end{subequations}
where $\delta$ is the ratio of diffusion constants and the functions $f(u,v)$
and $g(u,v)$ define the dynamics of the activator $u(\tilde{{\bm x}},\tilde{t})$ and
inhibitor $v(\tilde{{\bm x}},\tilde{t})$, respectively.
In the following we will consider two prototypical examples of activator-inhibitor 
dynamics widely analysed in the literature: the FitzHugh-Nagumo and  Brusselator models.
In both cases we will proceed as follows:
First we will study their fixed points and analyse their linear stability. 
The next step will be the derivation of the coefficients
 $c_1$ and $c_3$ of the CGLE (\ref{eq:cgle}).  
They can be used to decide the question whether spirals or antispirals 
appear near the Hopf bifurcation.
Finally we will compare these prediction with numerical simulations 
of both models near and far from onset.
%

\subsection{FitzHugh-Nagumo model}
The FitzHugh-Nagumo (FHN) dynamics \cite{FHN} is defined by \cite{Winfree}
\begin{subequations}
 \label{eq:FHN-dyn}
  \begin{eqnarray}
     f(u,v) &=& u -\frac{u^3}{3} - v ~,              \label{subeq:FHN-dyn-u}\\
     g(u,v) &=& \varepsilon ( u - \gamma v +\beta) ~,\label{subeq:FHN-dyn-v}
  \end{eqnarray}
\end{subequations}
where $\varepsilon>0$ is the ratio between the time scales of both fields,
and $\beta$ and $\gamma$ are parameters that determine the number of 
fixed points. 
The coordinates of these fixed points are independent of $\varepsilon$ and 
are given by the roots of a cubic polynomial. 
In the following we will only consider $\gamma>0$ and
restrict our analysis to the case where a unique fixed point $(u_0,v_0)$ exists.
This is the case, for example, for $\gamma=\frac{1}{2}$ where the coordinates of the fixed point are $u_0=({(\sqrt{1+9 \beta^2}-3\beta)^\frac{2}{3}-1})/{(\sqrt{1+9
\beta^2}-3\beta)^\frac{1}{3}}$ and $v_0=({u_0+\beta})/{\gamma}$.
Choosing $\varepsilon$ as the control parameter and keeping $\beta$, $\delta$ 
and $\gamma$ constant, a linear stability analysis of 
$(u_0,v_0)$ shows that it is unstable to periodic oscillations if  
$\varepsilon < \varepsilon_H^c$, where $\varepsilon_H^c=({1-u_0^2})/{\gamma}$.
This Hopf bifurcation has a frequency
$\Omega=\sqrt{(1-u_0^2)/\gamma-(1-u_0^2)^2}$ at onset.
The fixed point may also become unstable to spatially periodic perturbations 
with wavenumber $\tilde{q}_T=\sqrt{((1-u_0^2)\delta-\gamma \varepsilon)/(2\delta)}$ 
({\it i.e.} a Turing instability) if $\varepsilon < \varepsilon_T^c$, where 
$\varepsilon_T^c={((2-(1-u_0^2)\gamma+2 \sqrt{1-(1-u_0^2)\gamma})\delta)}/{\gamma^2}$.
%
\begin{figure*}
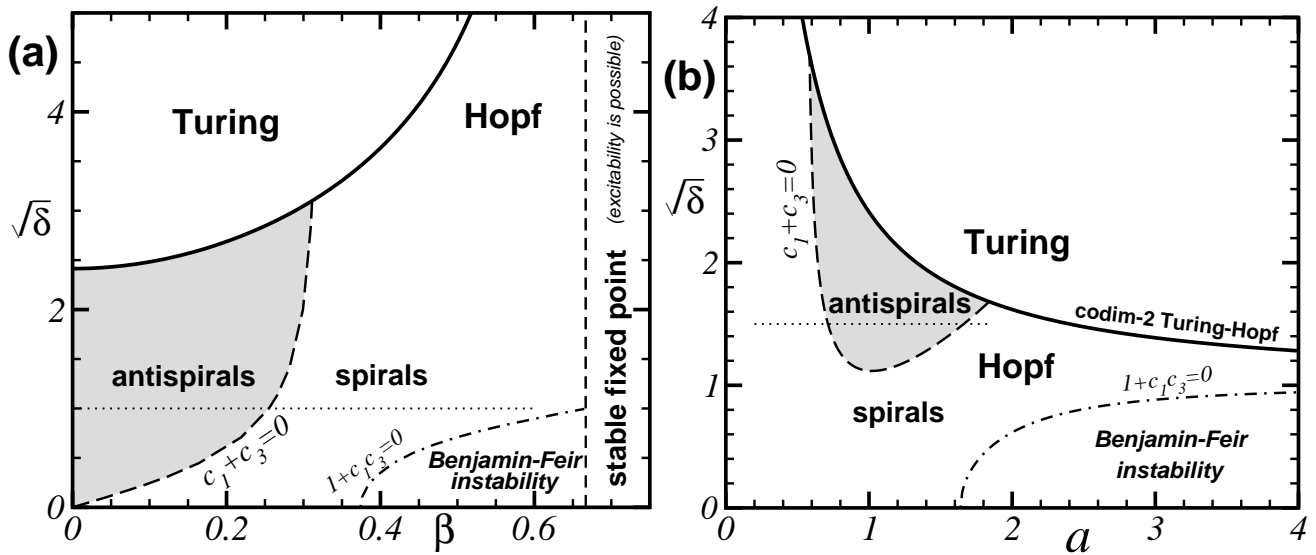

  \begin{center}
    \epsfxsize=0.48\hsize\mbox{\epsffile{figure3a.eps}}
    \epsfxsize=0.48\hsize \mbox{\epsffile{figure3b.eps}}
  \end{center}
\caption{
The gray areas show the  parameter space regions where antispirals are 
predicted to exist near the Hopf instability threshold for the FHN and  
Brusselator models (in (a) and (b), respectively). 
The thick full lines separates the regions where either  the Turing or Hopf
instability appears first as the control parameter $\varepsilon$ (parameter $b$) is decreased (increased) away from threshold for the FHN model in (a) (respectively, in (b) the  Brusselator model).
Inside the dot-dashed lines the Benjamin-Feir instability is predicted. 
The thin dotted lines indicate the range of  parameter values considered in
the numerical simulations of (anti)spirals summarised in 
Figs.~\ref{fig:spiral-antispiral-num-sym-FHN} and ~\ref{fig:spiral-antispiral-num-sym-Brusselator}.
The Hopf bifurcation considered in both figures is {\it supercritical}  (see main text).
In (a) we consider the parameter space $(\beta,\sqrt{\delta})$ of
the FHN model (Eqs. (\ref{eq:FHN-dyn})) with $\gamma=\frac{1}{2}$.
For $\beta\ge2/3$ the fixed point $(u_0,v_0)$ is stable and, if $\varepsilon$ is small
enough, excitability is possible.
In (b) the control parameter $(a,\sqrt{\delta})$ of Eq. (\ref{eq:Bruss-dyn}) is plotted.
}
  \label{fig:instability in parameter spaces}
\end{figure*}
%

These two instabilities may occur simultaneously at a codimension-2 instability.
The thick full line in Fig.~\ref{fig:instability in parameter spaces}(a) shows the 
location of this codimension-2 instability in the  parameter space 
$(\beta,\sqrt{\delta})$ for $\gamma=\frac{1}{2}$.
In the following we will restrict our analysis to the Hopf instability ({\it i.e}
the region below the codimension-2 line in Fig.~\ref{fig:instability in parameter spaces}(a)).
Using the methods explained in \cite{Nicolis,Kuramoto}, the derivation of the 
coefficients $c_1$ and $c_3$ is straightforward. 
Here we will only quote the main results of this derivation. 
The linear dispersion coefficient is given by
\begin{equation}
   c_1=\Big(\frac{1-u_0^2}{\Omega}\Big)\frac{\delta-1}{\delta+1} ~. \nonumber
\end{equation}
Note that this coefficient vanishes if both fields diffuse with equal strength.
If $({1-\gamma-\gamma u_{0}^2})/({1-\gamma+\gamma u_{0}^2})>0$ the Hopf bifurcation
is supercritical, otherwise it is subcritical and 
the CGLE (\ref{eq:cgle}) can not be applied.
For $\gamma=\frac{1}{2}$ (the case shown in Fig.~\ref{fig:instability in parameter spaces}(a))
the Hopf instability is supercritical for any value of $\beta$.
The result for the nonlinear dispersion coefficient $c_3$ is more complicated:
\begin{eqnarray}
c_3=\frac{3 - 3 \gamma  - 7 u_0^2 + 3 \gamma u_0^4}
         {3 - 3 \gamma  - 3 \gamma u_0^2  }
    \sqrt{ \frac  {\gamma } {\left( 1 - u_0^2 \right)  \left( 1 - \gamma  + 
    \gamma u_0^2 \right) } } ~.
    \nonumber
\end{eqnarray}
We may now use the expression for the coefficients $c_1$ and $c_3$ to evaluate the conditions for the 
spiral-antispiral transition $c_1+c_3=0$ and the Benjamin-Feir instability $1+c_1c_3=0$ in the parameter space  $(\beta,\sqrt{\delta})$.
These two lines are plotted in Fig.~\ref{fig:instability in parameter spaces}(a).
Note that the region where antispirals are predicted to occur near threshold, exists only for small values of $\beta$ ({\it i.e.} far from the excitability region) and vanishes if 
$\delta$ is too big or if $\delta \to 0$.
Moreover, for $\beta=0$, where the model exhibits the symmetry 
$(u,v)\to(-u,-v)$, only antispirals are predicted. 

The predicted spiral-antispiral transition at $c_1+c_3=0$ is strictly valid
only close to supercritical Hopf bifurcations ({\it i.e.} 
for $\varepsilon \lessapprox \varepsilon_H^c$).
In order to test this and in addition to investigate the behaviour far from  threshold,
we have studied the FHN model numerically. 
We have done extensive numerical simulations of 1D-(anti)spiral
analogues in Eq. (\ref{eq:FHN-dyn}) near and far from threshold.
These simulations were performed in very long systems (typically including hundreds of 
spiral wavelengths) with the following boundary conditions: 
in the right side a zero flux condition is imposed and 
in the left side the field is kept to $u=u_0$ and $v=v_0$.
%
\begin{figure*}
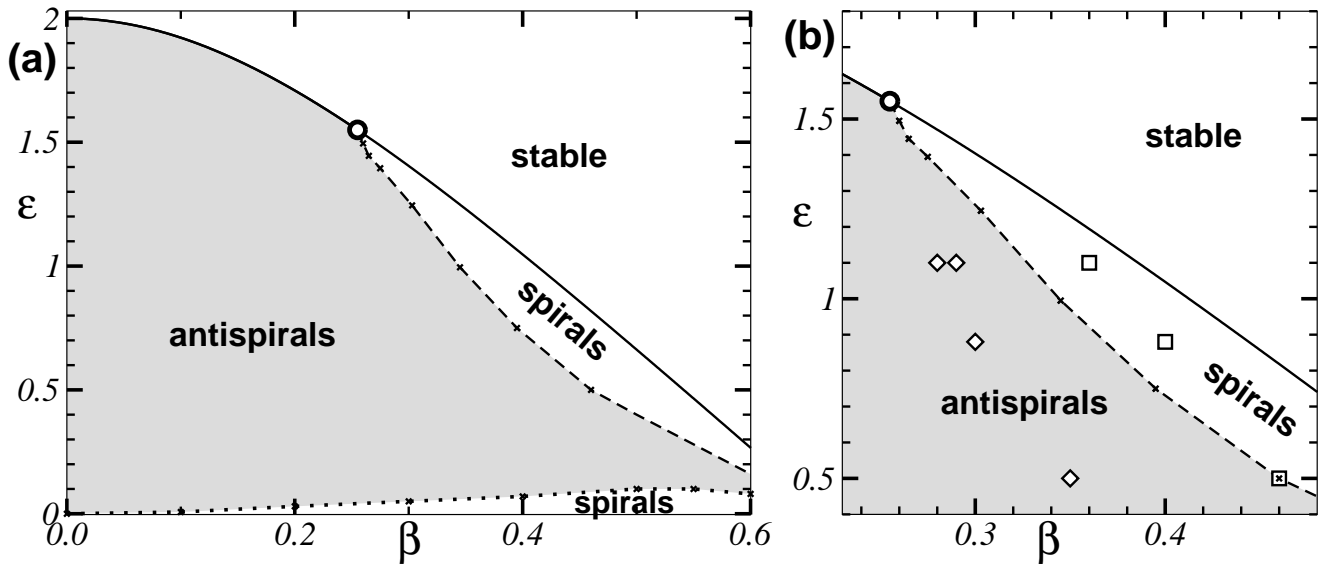

  \begin{center}
\epsfxsize=0.57\hsize\epsffile{figure4a.eps}
\epsfxsize=0.4\hsize\epsffile{figure4b.eps}
   \end{center}
\caption{
Summary of the results of numerical simulations of 1D- and 2D-(anti)spirals in the FHN  model.
The thick full line in (a) indicates the location of the Hopf instability threshold.
The white circle indicates the spiral-antispiral transition point at onset,
as predicted by the the criterion $c_1+c_3=0$ ({\it cf.} Fig.~\ref{fig:instability in parameter spaces}).
The dashed curve indicates the spiral-antispiral transition of the 1D-(anti)spiral analog which was computed numerically. 
The thick dotted line showd the antispiral-spiral transition occurring for small values of $\varepsilon$.
In (b) a magnification of the region where the spiral-antispiral transition occurs near onset is shown.
The squares and diamonds show the location of  2D numerical simulations ({\it cf.} Fig.~\ref{fig:snap}), resulting in spirals and antispiral respectively.
The parameters are $\sqrt{\delta}=1$ and $\gamma=\frac{1}{2}$.
Refer to Fig.~\ref{fig:instability in parameter spaces} for a resume of the
location of these simulations in the parameter spaces $(\beta,\sqrt{\delta})$.
}
\label{fig:spiral-antispiral-num-sym-FHN}
\end{figure*}
%
%
The output of these numerical simulations, see
Fig. \ref{fig:spiral-antispiral-num-sym-FHN}, is insensitive to initial conditions.
If we wait long enough, the boundary condition in the left side (the  ``core") will select phase
waves with a particular $\tilde{q}_{s1}$ which eventually invade the rest of the
system (since they always have positive $\tilde{v}_{gr}$).
Antispirals (spirals) will have negative (positive) phase velocity $\tilde{v}_{ph}$.
Near the Hopf threshold the observed behaviour is equivalent to the
one predicted by the CGLE. 
But, as $\varepsilon$ is decreased, a transition from antispirals to spirals is seen. 
This transition is related to a change in the sign of $\tilde{q}_{s1}$
and is not captured by the CGLE description (see below).
In order to assess the previous results for the 1D-(anti)spiral analog
we have also carried out a number of 2D simulations of the FHN model near onset. 
The outcome of these simulations is shown in
Fig.~\ref{fig:spiral-antispiral-num-sym-FHN}(b) and confirms that, at least near the onset, the 
1D analog provides a good description of the spiral-antispiral transition mechanism in the 2D system.
%
%
\begin{figure}
  \begin{center}
    \epsfxsize=\hsize \mbox{\epsffile{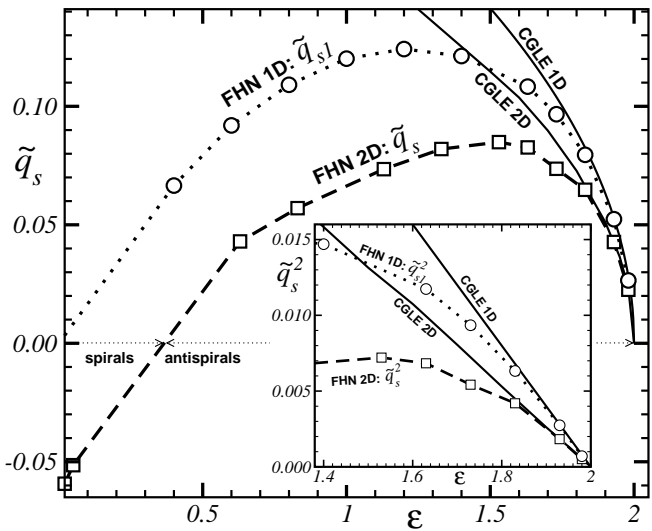}}
  \end{center}
\caption{
Plot of the measured wavenumber $\tilde{q}_s$ selected by 1D- and
2D-(anti)spirals (open circles and squares, respectively) in 
numerical simulations of the FHN model (\ref{eq:FHN-dyn}) as a
function  of the parameter $\varepsilon$, with $\sqrt{\delta}=1$ and  $\beta=0$ 
(in this case $\gamma=1/2$ and $\varepsilon_H^c = 2$).
The full lines indicate the wavenumber selected by the 2D antispirals in
and 1D antispiral analogues in the CGLE.
The {\bf CGLE 1D} line is calculated using the wavenumber 
$\tilde{q}_{s1}^2$ predicted analytically by Hagan's formula 
(\ref{eq:qhagan1D}). 
In the inset the value of $\tilde{q}_S^2$ in the region close to the onset is shown.
Note that the {\bf CGLE 1D} and {\bf CGLE 2D}  lines are directly
proportional to the distance to the threshold $\sqrt{\varepsilon_H^c
-\varepsilon}$ and consequently that both $q_{s1}$ and $q_s$ are
constant in the framework of the amplitude equations, but also note 
that this constant value differs \cite{q1D}.
The measured $\tilde{q}_{s1}$ and $\tilde{q}_{s}$ approach (as
expected) the predictions of the 1D and 2D CGLE
antispirals near the threshold but are systematically smaller far away
from it. 
}
\label{fig:fhn-ks-vs-eps}
\end{figure}
%
To see this more quantitatively, in Fig.~\ref{fig:fhn-ks-vs-eps} we
plot, against $\varepsilon$, the selected wavenumbers $\tilde{q}_s$,
$\tilde{q}_{s1}$, in 1D and 2D, and the corresponding wavenumbers in the CGLE 
in 1D and 2D (using the coefficients derived in this section), 
for the case $\beta=0$ and $\delta=1$.
Note, in the inset, the dramatic disagreement between the measured
wavenumbers and the predicted by the CGLE as the distance to the threshold is increased.
This figure makes explicit the change of sign of $\tilde{q}_s$ occurring for small values of $\varepsilon$.
Also note that, although the measured $\tilde{q}_s$ gets apart from
$\tilde{q}_{s1}$ as the distance to onset increases, the qualitative behaviour is the same.
%

\subsection{Brusselator model}

Let us finally address the Brusselator dynamics \cite{Brusselator}.
This model is defined by:
\begin{subequations}
 \label{eq:Bruss-dyn}
  \begin{eqnarray}
     f(u,v) &=& a- (b+1) u + u^2 v ~, \label{subeq:Bruss-dyn-u}\\
     g(u,v) &=& b u - u^2 v ~,        \label{subeq:Bruss-dyn-v}
  \end{eqnarray}
\end{subequations}
where $a$ and $b$ are two parameters (assumed in the following to be positive).
This model has a unique fixed point $(u_0,v_0)=(a,{b}/{a})$, for any value of the  
parameters $a$, $b$ and $\delta$.
Similarly to the FitzHugh-Nagumo model, we may now examine the stability of the 
fixed point.
Let us take $b$ as the control parameter.
Depending on the values of the parameters $a$ and $\delta$, the fixed point $(u_0,v_0)$
may become unstable in two different ways.
%
If $b>b_H^c$, where $b_H^c=1+a^2$, then a Hopf instability with frequency
$\Omega=a$ occurs.
%
In the other hand, if $b>b_T^c$, where $b_T^c=(1+{a}/{\sqrt{\delta}})^2$, a Turing 
instability with wavenumber $\tilde{q}_T^2={a/\sqrt{\delta}}$ takes place.
These instabilities may occur simultaneously in a codimension-2 line.
This line and the regions in the parameter space $(a,\sqrt{\delta})$ where either 
the Hopf or  Turing instabilities  appear first, as the control parameter $b$ 
is increased, are shown in Fig.~\ref{fig:instability in parameter spaces}(b). 

A short calculation \cite{Kuramoto} shows that the coefficients $c_1$ and $c_3$ are
\begin{equation}
c_1=a\frac{\delta-1}{\delta+1} \quad\mbox{and}\quad c_3=\frac{7a^2-4-4a^4}{3a(2+a^2)} ~.
\nonumber
\end{equation}
The spiral/antispiral transition line $c_1+c_3=0$ is plotted in the parameter
space $(a,\sqrt{\delta})$ in  Fig.~\ref{fig:instability in parameter spaces}(b).
Note that the region where antispirals are predicted is rather small
and is located near the area where of the Turing instability appears first.
Also in this case the antispirals will occur only for intermediate
values of the diffusion ratio $\delta$.
We have performed numerical simulations also for this model.
As for the FHN model, we investigated 1D-spirals and antispirals 
(with the same boundary conditions as in the previous study).
%
%
\begin{figure}
  \begin{center}  \epsfxsize=\hsize\mbox{\epsffile{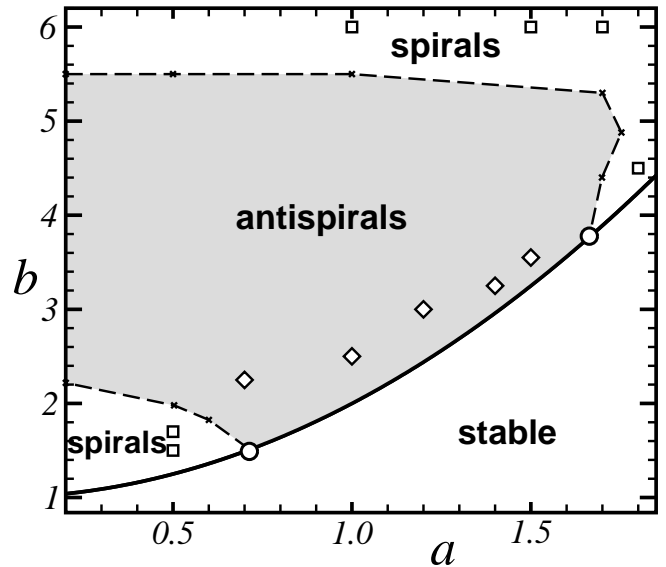}}
  \end{center}
\caption{
Summary of the results of numerical simulations of 1D- and
2D-(anti)spirals in the  Brusselator model with $\sqrt{\delta}=1.5$.
The thick full line indicates the location of the Hopf instability threshold and
the white circles indicate the spiral-antispiral transition points at onset
({\it cf.} Fig.~\ref{fig:instability in parameter spaces}).
The dashed curves indicate (approximately) the analogue of the spiral-antispiral
transition in the 1D calculations. 
The squares and diamonds show the location of some 2D numerical
simulations resulting in spirals and antispirals respectively.
}
\label{fig:spiral-antispiral-num-sym-Brusselator}
\end{figure}
%
The results of these simulations are resumed in 
Fig.~\ref{fig:spiral-antispiral-num-sym-Brusselator}.
In this figure we also show the output of some 2D simulations, which
confirm the observation that 1D simulations provide a good estimate
for the location of the spiral-antispiral transition.
Also here it is seen that the prediction based in the CGLE description
is valid only in the vicinity of the Hopf threshold. 
As the distance to this threshold increases the antispirals are turned into
spirals.
%

\section{Summary}

In this paper we emphasise the fact that the essential difference
between spirals and antispirals is the sign of phase velocity of the
travelling waves emanating from the core region and
that phase and group velocity do not necessarily have to  point into the same direction.
Consequently, we may conclude that spirals and antispirals can not occur simultaneously.
They appear in different regions of the parameter-space.
We have shown that spirals and antispirals are related by a simple symmetry 
transformation in the parameter space of the CGLE and  
derived a criterion for the CGLE coefficients, namely $c_1+c_3>0$,
that predicts the existence of antispirals in the respective RD model near a Hopf instability.
In the CGLE parameter space, antispirals appears as frequently as spirals. 
Spirals and antispirals are found in complementary regions of the
paramter space and are characterized by the selection of positive and
negative wavenumbers, respectively. 
At the boundary between antispirals and spirals, the selected
wavenumber is zero and the selected wavelength diverges. 
Our criterion has been mapped to the parameter space of two
 representative reaction-diffusion models.
The analysis of both models suggests that to get antispirals it is 
necessary that the inhibitor also diffuses.
Moreover, the diffusion constants of both reactants should be of
similar magnitude in order to find antispirals, which
is likely to occur near the codimension-2 Turing-Hopf instability.
We have also performed numerical simulations of both models 
near and far form the oscillatory instability threshold.
Antispirals are quite likely to appear near the Hopf bifurcation
but turn into regular spirals far into the oscillatory
where the amplitude of waves becomes large.
Both examples show that the assumption that the 1D spiral
analog gives a good approximation of the spiral-antispiral transition
behaviour not only near onset but also moderately far from it. 

In a more general distinction, antispirals emit phase waves that are 
typical for  oscillatory media. 
In contrast, trigger waves common in
excitable media should always have the same sign of group and phase
velocity. 
Since the distinction between trigger and phase waves is not sharp, we
cannot make a general statement about the possibility of antispirals
in excitable media, but expect that they will be quite hard to find
therein. 
The rarity of experimental observations of antispirals suggests that
most experiments exhibiting spirals are conducted either far away from the oscillatory
threshold or under excitable resp. bistable conditions. 
It is tempting to explain the experiment of Vanag and Epstein
\cite{Vanag} from the proximity of the control parameters to a Hopf
bifurcation, though other more complicated possibilities cannot be ruled out at this
stage.

Finally, we like to stress the most essential point in the consideration of
the nature of spirals and antispirals. 
The crucial quantity for both structures is the group velocity of the
periodic waves outside the core region. 
In all models considered, the group velocity is pointing outward from
the spiral core as has been shown analytically for the CGLE. 
The distinction between spirals and antispirals stems only from the 
sign of the phase velocities and the corresponding phenomenological
impression. 
In our opinion is it important not to confuse the distinction between
spirals and antispirals with the distinction between sinks and
sources. 
Several authors have pointed out that the definition of a source depends
only on the group velocities and not at all on the phase velocity of
the periodic waves far away from it  \cite{MvH,sandstede}.
Counting arguments show that sources belong typically to a
zero-parameter family while sinks are members of a two parameter 
family \cite{MvH}. 
In other words, sources select frequency and wavelength of the emitted
periodic waves, while sinks simply appear where two wavetrains of
arbitrary wavelength and period happen to meet. 
A classification of defects in reaction-diffusion media can be done
according to the group velocities of periodic waves on both sides of
the defect and the velocity of the defect itself.
Apart from sources and sinks, one may also find transmission defects
(one-parameter family) and contact defects in reaction-diffusion media
\cite{sandstede}.  
In the present context, spirals and antispirals are both sources,
i. e. they select their frequency and wavelength and control their
surroundings.
If they were sinks (as sometimes claimed in the literature), the waves
had to be organized from the domain boundaries, which would require
another unknown ``organizer'' structure. 
Altogether, we have shown by combination of existing analytical
arguments that antispirals are structures of similar nature as spirals
and should hence be expected in many systems exhibiting phase waves in
particular near the onset of oscillations. 
Their experimental discovery can be nicely interpreted within
the existing theoretical framework for oscillatory reaction-diffusion media.
%

\acknowledgments
We are grateful to I. Aranson, L. Kramer, H.~Swinney and V. Vanag for discussions. 
Financial support by the {\it German-Israeli Foundation (GIF)} is gratefully
acknowledged. 
E.~M.~Nicola would like to thank financial support given by the 
{\it Sonderforschungsbereiche (SFB) 555 der DFG: ``Komplexe nichtlineare 
Prozesse" (Berlin, Germany)} and by the European Commission under network HPRN-CT-2002-00312.
L.~Brusch  acknowledges the support of the {\em Max Planck Society} through an 
Otto Hahn fellowship.
%



\begin{thebibliography}{99}
%

\bibitem{Showalter}
R. Kapral, and K. Showalter (Eds.),
{\it Chemical waves and patterns},
(Kluwer, Dordrecht, 1994).


\bibitem{ImbErtl}
R. Imbihl and G. Ertl, Chem. Rev. {\bf 95}, 697 (1995).

\bibitem{Murray}
J. Murray, {\it Mathematical Biology},
(Springer, Berlin, 1989).

\bibitem{keener} 
J. Keener and J. Sneyd,
{\it Mathematical Physiology},
(Springer, New York, 1998).




\bibitem{winfree72} 
A. T. Winfree, Science {\bf 175}, 634 (1972).

\bibitem{dicty}
G. Gerisch, Naturwissenschaften {\bf 58}, 430 (1971); 
P. Devreotes, Science {\bf 245}, 1045 (1989). 

\bibitem{catalysis} 
S. Jakubith, 
H.-H. Rotermund, W. Engel, A. von Oertzen, and G. Ertl, 
Phys. Rev. Lett. {\bf 65}, 3013 (1990); 
S. Nettesheim, A. von Oertzen, H.H. Rotermund, and G. Ertl,
J. Chem. Phys. {\bf 98}, 9977 (1993).

\bibitem{jalife} 
J. M. Davidenko, A. M. Pertsov, R. Salomonsz, W. Baxter and J. Jalife, 
Nature {\bf 353}, 349 (1991). 

\bibitem{lechl}
J. Lechleiter, S. Girard, E. Peralta and D. Clapham, 
Science {\bf 252}, 123 (1991) 

\bibitem{mair} 
Th. Mair and  S. C. M\"uller,
J. Biol. Chem.  {\bf 271}, 627 (1996). 





%
\bibitem{Vanag}
V.~K.~Vanag and I.~R.~Epstein, Science {\bf 294}, 835 (2001) 
%

%
\bibitem{YK76}
T.~Yamada and Y.~Kuramoto, Prog.~Theor.~Phys., {\bf 55}, 2035 (1976).
%
\bibitem{selpuchre} 
J. A. Selpuchre, G. Dewel, and A. Babloyantz, 
Phys. Lett. A {\bf 147}, 380 (1990). 

\bibitem{kapral} 
A.  Goryachev and R. Kapral, 
Phys. Rev. Lett. {\bf 76}, 1619 (1996).  

\bibitem{Ipsen97}
M. Ipsen M, F. Hynne and P. G. Sorensen, Int. J. Bif. and Chaos {\bf 7},
1539 (1997).

\bibitem{TK98}
S.M. Tobias and E. Knobloch, Phys.~Rev.~Lett. {\bf 80}, 4811 (1998).

\bibitem{Rabinovitch} 
A. Rabinovitch, M. Gutman, and I. Aviram, 
Phys. Rev. Lett. {\bf 87}, 084101 (2001).  

\bibitem{Stich}
M. Stich and A. Mikhailov, Z.~Phys.~Chem. {\bf 216}, 521 (2002).

\bibitem{Lee}
S.-J. Woo, J. Lee and K. J. Lee, Phys. Rev. E {\bf 68}, 016208
(2003). 

\bibitem{soerensen}
H.~Sk{\o}dt and P.~G.~S{\o}rensen, Phys.~Rev.~E {\bf 68}, 020902 (2003).

\bibitem{review} 
I.~S.~Aranson and L.~Kramer, Rev.~Mod.~Phys. {\bf 74}, 99 (2002). 
%
\bibitem{Vanag3}
V.~K.~Vanag and I.~R.~Epstein,
Phys.~Rev.~Lett. {\bf 88}, 088303 (2002). 
%

\bibitem{Nicola} 
E. M. Nicola, W. Wolf, M. Or-Guil and M. B\"ar, 
 Phys.~Rev.~E {\bf 65}, 055101 (2002).


\bibitem{christini}
Y.~Gong and D.~J.~Christini, 
Phys.~Rev.~Lett. {\bf 90}, 088302 (2003). 
%

\bibitem{BNB03}
L. Brusch, E. M. Nicola and M. B\"ar, 
Phys. Rev. Lett. {\bf 92}, 089801 (2004).
%


%
\bibitem{Kuramoto}
Y.~Kuramoto, {\it Chemical Oscillations, Waves, and Turbulence},
(Springer-Verlag, Berlin, 1984).
%

%
\bibitem{CH}
M.~C.~Cross and P.~C.~Hohenberg, Rev.~Mod.~Phys. {\bf 65}, 851 (1993).
%

%
\bibitem{Nicolis} 
G.~Nicolis, {\it Introduction to Nonlinear Sciences},  
(Cambridge University Press, Cambridge, 1995).
%

%

\bibitem{Ipsen00}
M. Ipsen, L. Kramer, P. G. S{\o}rensen, Phys. Rep. {\bf 337}, 193 (2000).






%
\bibitem{KT76}
Y.~Kuramoto and T.~Tsuzuki, Prog.~Theor.~Phys. {\bf 55}, 356 (1976).
%


\bibitem{Vanag2}
V.~K.~Vanag and I.~R.~Epstein,
Phys.~Rev.~Lett. {\bf 87}, 228301 (2001). 
%



\bibitem{Yang}
L.~Yang, M.~Dolik, A.~M.~Zhaotinsky and I.~R.~Epstein
J.~Chem.~Phys. {\bf 117}, 7258 (2002).

%
\bibitem{Brusselator}
I.~Prigogine, and R.~Lefever, J.~Chem.~Phys. {\bf 48}, 1695 (1968).
%

%
\bibitem{FHN}
R.~FitzHugh, Biophys. {\bf 1}, 445 (1961);
J.~S.~Nagumo., S.~Arimoto \and S.~Yoshizawa, Proc.~IRE {\bf 50}, 2061 (1962).
%


%
\bibitem{Hagan} 
P.~S.~Hagan, SIAM~J.~Appl.~Math. {\bf 42}, 762 (1982). 
%

%
\bibitem{anti1}
S.~Komineas, F.~Heilmann and L.~Kramer, Phys.~Rev.~E {\bf 63}, 011103 (2000).
%

%
\bibitem{q1D}
E.~Bodenschatz, A.~Weber and L.~Kramer, 
in {\it Nonlinear Wave Processes in Excitable Media}, 
edited by  A.~V.~Holden, M.~Markus and H.~G.~Othmer 
(Plenum Press, New York, 1990);
I.~S. Aranson~{\it et~al.}, Phys.~Rev.~A {\bf 46}, 2992 (1992). 
%
%



%
\bibitem{Ermentrout}
J.~Paullet, B.~Ermentrout and W.~Troy, SIAM~J.~Appl.~Math. {\bf 54}, 1386 (1994).
%



%
\bibitem{Winfree}
A.~T.~Winfree, Chaos {\bf 1}, 303 (1991).
%
\bibitem{MvH}
M.~van~Hecke, C.~Storm and W.~van~Saarloos, Physica~D {\bf 134},1 (1999). 
%
\bibitem{sandstede} 
B. Sandstede and A. Scheel, SIAM J. Applied Dynamical Systems, in press
(2004). 

\end{thebibliography}
\end{document}